\documentclass[epsfig]{article}
\usepackage{epsfig}
\usepackage{amssymb}
\catcode`\@=11
\def\@normalsize{\@setsize\normalsize{15pt}\xiipt\@xiipt
\abovedisplayskip 14pt plus3pt minus3pt%
\belowdisplayskip \abovedisplayskip
\abovedisplayshortskip  \z@ plus3pt%
\belowdisplayshortskip  7pt plus3.5pt minus0pt}
\def\small{\@setsize\small{13.6pt}\xipt\@xipt
\abovedisplayskip 13pt plus3pt minus3pt%
\belowdisplayskip \abovedisplayskip
\abovedisplayshortskip  \z@ plus3pt%
\belowdisplayshortskip  7pt plus3.5pt minus0pt
\def\@listi{\parsep 4.5pt plus 2pt minus 1pt
            \itemsep \parsep
            \topsep 9pt plus 3pt minus 3pt}}
\def\underline#1{\relax\ifmmode\@@underline#1\else
        $\@@underline{\hbox{#1}}$\relax\fi}
\@twosidetrue \relax \catcode`@=12
\catcode`\@=11

\def\section{\@startsection{section}{1}{\z@}{3.5ex plus 1ex minus
   .2ex}{2.3ex plus .2ex}{\large\bf}}

\def\ps@headings{\def\@oddfoot{}\def\@evenfoot{}
\def\@oddhead{\hbox{}\hfill
        \makebox[.5\textwidth]{\raggedright\ignorespaces --\thepage{}--
        \hfill }}
\def\@evenhead{\@oddhead}
\def\subsectionmark##1{\markboth{##1}{}}
} \ps@headings \catcode`\@=12 \relax

\def\figcap{\section*{Figure Captions\markboth
        {FIGURECAPTIONS}{FIGURECAPTIONS}}\list
        {Fig. \arabic{enumi}:\hfill}{\settowidth\labelwidth{Fig. 999:}
        \leftmargin\labelwidth
        \advance\leftmargin\labelsep\usecounter{enumi}}}
 \relax
\def\tablecap{\section*{Table Captions\markboth
        {TABLECAPTIONS}{TABLECAPTIONS}}\list
        {Table \arabic{enumi}:\hfill}{\settowidth\labelwidth{Table 999:}
        \leftmargin\labelwidth
        \advance\leftmargin\labelsep\usecounter{enumi}}}
 \relax
\def\reflist{\section*{References\markboth
        {REFLIST}{REFLIST}}\list
        {[\arabic{enumi}]\hfill}{\settowidth\labelwidth{[999]}
        \leftmargin\labelwidth
        \advance\leftmargin\labelsep\usecounter{enumi}}}
 \relax
\catcode`\@=11
\def\marginnote#1{}
\newcount\hour
\newcount\minute
\newtoks\amorpm
\hour=\time\divide\hour by60 \minute=\time{\multiply\hour by60
\global\advance\minute by- \hour}
\edef\standardtime{{\ifnum\hour<12 \global\amorpm={am}%
    \else\global\amorpm={pm}\advance\hour by-12 \fi
    \ifnum\hour=0 \hour=12 \fi
    \number\hour:\ifnum\minute<100\fi\number\minute\the\amorpm}}
\edef\militarytime{\number\hour:\ifnum\minute<100\fi\number\minute}

\def\draftlabel#1{{\@bsphack\if@filesw {\let\thepage\relax
  \xdef\@gtempa{\write\@auxout{\string
    \newlabel{#1}{{\@currentlabel}{\thepage}}}}}\@gtempa
    \if@nobreak \ifvmode\nobreak\fi\fi\fi\@esphack}
     \gdef\@eqnlabel{#1}}
\def\@eqnlabel{}
\def\@vacuum{}
\def\draftmarginnote#1{\marginpar{\raggedright\scriptsize\tt#1}}
\def\draft{\oddsidemargin -.5truein
        \def\@oddfoot{\sl preliminary draft \hfil
        \rm\thepage\hfil\sl\today\quad\militarytime}
        \let\@evenfoot\@oddfoot \overfullrule 3pt
        \let\label=\draftlabel
        \let\marginnote=\draftmarginnote
\def\@eqnnum{(\theequation)\rlap{\kern\marginparsep\tt\@eqnlabel}%
\global\let\@eqnlabel\@vacuum}  }
\def\preprint{\twocolumn\sloppy\flushbottom\parindent 1em
        \leftmargini 2em\leftmarginv .5em\leftmarginvi .5em
        \oddsidemargin -.5in    \evensidemargin -.5in
        \columnsep 15mm \footheight 0pt
        \textwidth 250mmin      \topmargin  -.4in
        \headheight 12pt \topskip .4in
        \textheight 175mm
        \footskip 0pt
\def\@oddhead{\thepage\hfil\addtocounter{page}{1}\thepage}
        \let\@evenhead\@oddhead \def\@oddfoot{} \def\@evenfoot{}
}
\def\titlepage{\@restonecolfalse\if@twocolumn\@restonecoltrue\onecolumn
     \else \newpage \fi \thispagestyle{empty}\c@page\z@
        \def\thefootnote{\fnsymbol{footnote}} }
\def\endtitlepage{\if@restonecol\twocolumn \else  \fi
        \def\thefootnote{\arabic{footnote}}
        \setcounter{footnote}{0}}  
\catcode`@=12 \relax

\def\ps@headings{\def\@oddfoot{}\def\@evenfoot{}
\def\@oddhead{\hbox{}\hfill
        \makebox[.5\textwidth]{\raggedright\ignorespaces --\thepage{}--
        \hfill }}
\def\@evenhead{\@oddhead}
\def\subsectionmark##1{\markboth{##1}{}}
} \ps@headings \relax
\newcommand{\newc}{\newcommand}

\newc{\ra}{\rightarrow}
\newc{\lra}{\leftrightarrow}
\newc{\beq}{\begin{equation}}
\newc{\be}{\begin{equation}}
\newc{\eeq}{\end{equation}}
\newc{\ee}{\end{equation}}
\newc{\bea}{\begin{eqnarray}}
\newc{\eea}{\end{eqnarray}}

\newc{\ome}{\omega}
\newc{\ba}{\begin{eqnarray}}
 \newc{\ea}{\end{eqnarray}}

\begin{document}
\def\firstpage#1#2#3#4#5#6{
\begin{titlepage}
\nopagebreak
\title{\begin{flushright}
        \vspace*{-0.8in}
{ 
}
\end{flushright}
\vfill {#3}}
\author{\large #4 \\[1.0cm] #5}
\maketitle \vskip -7mm \nopagebreak
\begin{abstract}
{\noindent #6}
\end{abstract}
\vfill
\begin{flushleft}

February 2008
\end{flushleft}
\thispagestyle{empty}
\end{titlepage}}

\def\simlt{\stackrel{<}{{}_\sim}}
\def\simgt{\stackrel{>}{{}_\sim}}
\date{}
\firstpage{3118}{IC/95/34} {\large\bf A $U(3)_{C}\times
U(3)_{L}\times U(3)_{R}$ gauge symmetry from intersecting D-branes}
 {G.K. Leontaris}
{\normalsize\sl Theoretical Physics Division, Ioannina University,
GR-45110 Ioannina, Greece\\
\\ [2.5mm]
 }
{A three-family non-supersymmetric model with $U(3)^3$ gauge
symmetry is analyzed in the context of intersecting D-branes. This
is equivalent to the `trinification' model extended by three $U(1)$
factors which survive as global symmetries in the low energy
effective model. The Standard Model fermions are accommodated in the
three possible bifundamental multiplets represented by strings with
endpoints attached on different brane-stacks of this particular
setup.  Further, a natural Quark-Lepton hierarchy is realized  due
to the existence of the additional abelian symmetries. }

\vskip 3truecm

\newpage

 The ascertainment that the non-perturbative sector of string theory
 includes higher dimensional objects~\cite{Polchinski:1995mt}, the
 so called Dp-branes, opened up new horizons in model building.
 The subsequent observation that it is possible to obtain chiral fermions
 from D-branes intersecting at angles~\cite{Berkooz:1996km},
 was of crucial importance towards this direction. In these constructions,
 matter fields are represented by open strings stretched between
intersecting  D-brane stacks and they are localized in the
intersection locus of the latter. During the last few years, several
authors have constructed  semi-realistic D-brane models  with or
without supersymmetry~\footnote{For reviews see for
example~\cite{Kiritsis:2003mc},\cite{Blumenhagen:2006ci} and
references therein.}, anticipating that the unknown parameters of
the Standard Model (SM) (i.e., mass spectrum etc) will be determined
in terms of geometric quantities related to the size and shape of
the compactification
space\cite{Blumenhagen:2000wh}-\cite{Anastasopoulos:2006da}.

In a previous paper\cite{Leontaris:2005ax} we proposed a D-brane
inspired non-supersymmetric model with  $U(3)_C\times U(3)_L\times
U(3)_R$ gauge symmetry and explored in some detail several of its
low energy implications. It was found that the emerging effective
field theory model is identical with the so-called trinification
model proposed long time ago~\cite{Rizov:1981dp} and subsequently
analyzed in a superstring context~\cite{Ganoulis:1988vg}. In the
present work, we will examine possible realizations of this model in
the context of D6-branes intersecting at angles. We will restrict
our present endeavors in the non-supersymmetric case although a
supersymmetric version could be also elaborated in a similar manner,
provided that proper constraints are taken into account.\footnote{In
general, D-brane configurations intersecting at arbitrary angles
lead to non-supersymmetric models. These have the shortcomings of
possible instabilities due to uncanceled NSNS-tadpoles but we will
assume that they can be eliminated realizing the construction on an
appropriate orbifold (for a consideration of these issues see for
example \cite{Rabadan:2001mt}).}

As mentioned, a Dp-brane is an extended object in p-dimensions. In
D-brane model building one exploits the fact that a stack of $N$
parallel, almost coincident  D-branes gives rise to a $U(N)$ gauge
group. Chirality arises~\cite{Blumenhagen:2000wh} when intersecting
branes are wrapped on a torus with  chiral fermions sitting in the
various intersections of the configuration. Here, the
six-dimensional (6-d) compact space is taken to be a 6-d
factorizable torus $T^6=T^2\times T^2\times T^2$.
 To construct the D-brane analogue of the trinification model, we
consider three stacks of D6-branes, each stack containing 3 parallel
almost coincident branes giving rise to the gauge symmetry
 \ba
U(3)_C\times U(3)_L\times U(3)_R\label{d3sym}
\ea
The first $U(3)_C$ is identified with the  $SU(3)$ color group, the
second involves the weak $SU(2)_L$ gauge symmetry and the third is
related to the $SU(2)_R$ gauge group. Since $U(3)= SU(3)\times U(1)$
this D-brane setup, in addition to the standard trinification gauge
group,  comprises also three extra $U(1)$ gauged  symmetries, thus
the  gauge symmetry in (\ref{d3sym}) can be equivalently written
\ba
 SU(3)_C\times SU(3)_L\times SU(3)_R\times U(1)_C\times
U(1)_L\times U(1)_R\label{333111}
\ea
In the D-brane context, matter fields appear as open strings having
 their endpoints attached to some of the brane stacks. An open string
can have its ends attached on any of the three  brane-stacks and
upon introducing orientifold planes, string-ends may also be
attached on the mirror brane-sets. In our particular construction,
 the following possibilities arise with regard to the spectrum: (i)
 strings with endpoints attached on two different 3-brane stacks belonging
 to bifundamentals $(3,\bar 3)$ or $(\bar 3,3)$  of the corresponding
gauge group factors, (ii) strings with one end on the 3-brane stack
and the other on a mirror brane-stack giving rise to $(3,3)$ or
$(\bar 3,\bar 3)$ representations; (iii) strings with both
string-ends found on the same brane-stack, which implies the
possible existence of antisymmetric $3_a$ and/or symmetric $6_s$
representations of the corresponding $U(3)$ gauge factor.

 The  appropriate representations to accommodate the standard
model particles and Higgs fields are the bifundamental multiplets.
Under the symmetry `decomposition' (\ref{333111}) these lead to the
following matter representations
\ba
{\cal Q^{\hphantom{c}}}&=&(3,\bar 3,1)_{(+1,-1,\hphantom{+}0)}\label{QL}\\
{\cal Q}^c&=&(\bar 3,1,3)_{(-1,\hphantom{+}0,+1)}\label{QR}\\
{\cal L}^{\hphantom{c}}&=&(1,3,\bar
3)_{(\hphantom{+}0,+1,-1)}\label{LH}
\ea
Here, we adopt a notation where the three first numbers in
(\ref{QL}-\ref{LH}) refer to the color, left and right $SU(3)$ gauge
groups, while the three indices correspond to the three
$U(1)_{C,L,R}$ symmetries respectively. It turns out that these
three representations suffice to accommodate all the fermions of the
Standard Model. In particular, under the usual hypercharge embedding
\ba
 Y=-\frac{1}{6}
X_{L'}+\frac 13 X_{R'}\label{HC}
\ea
where $X_{L'}$ and $X_{R'}$ represent the $U(1)_{L'}$  and
$U(1)_{R'}$ generators in $SU(3)_{L}$  and $SU(3)_{R}$ respectively,
it can be observed that  the left-handed quark doublets $q=(u,d)^T$
and an additional colored triplet $D$ with SM-quantum numbers as
those of the down quark, can be identified in the representation
(\ref{QL}). Similarly,  representation (\ref{QR}) involves the
right-handed partners of (\ref{QL}). Finally (\ref{LH}) comprises
the lepton doublet, the right-handed electron and its corresponding
neutrino, two additional $SU(2)_L$ doublets and another neutral
state~\cite{Rizov:1981dp}. For a single family, the following
assignment is introduced
\ba
(3,\bar 3,1)\,=\,\left(\begin{array}{ccc}
u_r&d_r&D_r\\
u_g&d_g&D_g\\
u_b&d_b&D_b
\end{array}\right),\;
(\bar 3,1,3)\,=\,\left(\begin{array}{ccc}
u^c_r&u^c_g&u^c_b\\
d^c_r&d^c_g&d^c_b\\
D^c_r&D^c_g&D^c_b
\end{array}\right),\;
(1,3,\bar 3)\,=\,\left(\begin{array}{ccc}
E^{c0}&E^-&e\\
E^{c+}&E^0&\nu\\
e^c&\nu^{c+}&\nu^{c-}
\end{array}\right).\label{familas}
\ea
The Higgs multiplets responsible for the symmetry breaking down to
the Standard Model are accommodated in a representation identical to
that of the lepton fields,
\ba
{\cal H}_a &=&(1,3,\bar 3)_{(\hphantom{+}0,+1,-1)},\;\, a=1,2,...
\ea
Hence, the bifundamental representations suffice to accommodate all
fermion fields and the Higgs doublets of the Standard Model.
However, as mentioned previously, additional fields could arise from
strings with one of their ends attached on a mirror brane. These are
${\cal Q}^{'}=(3, 3,1)$, ${\cal Q}^{'c}=(3, 1,3)$, ${\cal L}^{'}=(1,
3,3)$ and their complex conjugates. In addition,
antisymmetric/symmetric representations arise from strings having
both their endpoints on the same brane stack. The quantum numbers of
the antisymmetric representations under the full gauge group are $
{\cal A}_{\cal C}=(3,1,1)_{(-2,0,0)}$,  ${\cal A}_{\cal
L}=(1,3,1)_{(0,-2,0)}$, ${\cal A}_{\cal R}=(1,1,3)_{(0,0,-2)}$ (plus
their complex conjugates) and similarly for the symmetric ones.
Notice however, that several of the above additional states under
the standard hypercharge definition, carry exotic (non-standard)
charges. For example, ${\cal A}_{{\cal L},{\cal R}}$ are
fractionally charged with charges $Q_{{\cal A}_{{\cal L},{\cal
R}}}=(\pm 1/3,\pm 2/3)$. The appearance of fractionally charges
states is a generic phenomenon in string theories and could be a
potential problem as long as  they remain in the light spectrum.
Interestingly, in the context of the D-brane analogue of the
trinification model, it is possible to remedy this problem by
redefining the hypercharge using an appropriate (anomaly-free)
linear combination of the extra abelian factors.

One of the characteristics of string derived models is the
appearance of anomalous $U(1)$ symmetries. In the present model all
three $U(1)_{C,L,R}$ appear to be anomalous, however, it has been
shown~\cite{Leontaris:2005ax} that there is an anomaly-free $U(1)$
combination
\ba
U(1)_{{\cal Z}'}&=& U(1)_C+U(1)_L+U(1)_R\label{aas}
\ea
The remaining  two orthogonal combinations $U(1)_C-U(1)_R$ and
$U(1)_C-2U(1)_L+U(1)_R$  have anomalies which are canceled   through
a generalized Green--Schwarz mechanism~\cite{Dine:1986zy} providing
masses to the corresponding  gauge bosons.  It can be checked that
the exotic  states are charged under (\ref{aas}), while the
bifundamentals accommodating the SM fermion and Higgs matter fields
are neutral under the anomaly-free $U(1)_{\cal Z}$. In this case it
possible to define a new hypercharge component
\ba
Y_{Z'}&=&Q_C+Q_L+Q_R
\ea
(where $Q_{C,L,R}$ refer to the $U(1)_{C,L,R}$ charges respectively)
corresponding to $U(1)_{\cal Z}$, so that the hypercharge definition
is generalized as follows
\ba
 Y'=-\frac{1}{6}
X_{L'}+\frac 13 X_{R'}-\frac 16 Y_{Z'}\label{NHC}
\ea
The previously fractionally charged states, now under the definition
(\ref{NHC}) obtain integral charges, while SM states being neutral
under $U(1)_{\cal Z}$ are unaffected.

In order to present a specific realization of the `trinification'
model in the intersecting D-brane context, we turn now to a short
description of the essential ingredients for model
building~\cite{Blumenhagen:2006ci}. In the intersecting D-brane
scenario, the multiplicities of the chiral fermions are related to
the open stings attached on the various brane-stacks, expressed in
terms of the various intersections of the
configuration~\cite{Berkooz:1996km,Blumenhagen:2000wh,Ibanez:2001nd}.
In particular the fermion generations belong to the bifundamentals
of the $D_a-D_b$ sectors of the theory, where $D_{a,b}$ stand for
the color ($D_c$), left ($D_l$) and right ($D_r$) 3-brane stacks.
Thus, we assign $D_c-D_l$ the sector giving origin to the
bifundamental $(3,\bar 3,1)$, while for the $D_c-D_r$ and $D_l-D_r$
sectors the relevant bifundamentals are $(\bar 3,1,3)$ and
$(1,3,\bar 3)$ respectively. The number of intersections for a
$D_a-D_b$ sector is given by
\ba
I_{ab}&=&\left(m_{a1} n_{b1}-m_{b1} n_{a1}\right)\left(m_{a2}
n_{b2}-m_{b2} n_{a2}\right)\left(m_{a3} n_{b3}-m_{b3} n_{a3}\right)
\ea
where the $(n_{ai},m_{ai})$ represent the winding numbers of the
$D_a$-stack around the two radii of the $i$-th torus. In the present
construction, we will allow for the possibility of magnetic fluxes
and let  $m_{ai}$ admit also half-integer values (tilted torii).
Additional matter fields arise from the sectors $D_a-\Omega\,{\cal
R}\,D_b$\footnote{ $\Omega$ is the world-sheet parity and ${\cal R}$
reflection operator. The combined action of the $\Omega\,{\cal R}$
symmetry means that for each $(n,m)$ wrapping pair there is also a
partner $(n,-m)$.}, while the relevant representations belong to
$(3, 3,1)$, $(3, 1,3)$ and $(1, 3,3)$ plus their conjugate fields.
Their multiplicity is given by
\ba
I_{ab^*}&=&-\left(m_{a1} n_{b1}+m_{b1} n_{a1}\right)\left(m_{a2}
n_{b2}+m_{b2} n_{a2}\right)\left(m_{a3} n_{b3}+m_{b3} n_{a3}\right)
\ea
Finally, for each gauge $U(3)$ group factor, there are in principle
symmetric ($6_s$) as well as antisymmetric ($3_a$) representations
due to the invariant $D_a-\Omega\,{\cal R}\,D_a$ sector. The three
antisymmetric representations are $(3,1,1)$, $(1,3,1)$, $(1,1,3)$
and similarly for the symmetric ones. Thus, the $D_a-\Omega\,{\cal
R}\,D_a$ sector provides
\ba
I_{aa^*}&=& 8\,m_{a1} m_{a2} m_{a3}\label{antis}
\ea
antisymmetric representations and
 \ba
 I'_{aa^*}&=&4(n_{a1} n_{a2}
n_{a3}-1) m_{a1} m_{a2} m_{a3}\label{symet}
 \ea
 symmetric as well as antisymmetric representations for
 the particular gauge factor represented by the $D_a$ brane-stack.

 The winding numbers  which determine the multiplicity of
 the spectrum and other crucial parameters (i.e. gauge and Yukawa couplings)
 of the effective field theory, are subject  to several  restrictions. The
 most important are the Tadpole  and anomaly cancelation conditions.
 Furthermore, conditions should also be imposed on  $n_{ai}, m_{ai}$
 in order to obtain the  correct number of fermion generations and Higgs
 fields, and at  the same time to eliminate unwanted states from the spectrum.
  Before we analyze in detail the conditions with respect to the spectrum of
  the particular model,  we summarize for convenience the Tadpole and anomaly
   cancelation conditions.

The  restrictions imposed  on the $(n_{ai}, m_{ai})$ sets
originating from the $RR$-type tadpole conditions read:
\ba
T_0\;=\;\sum_{a=c,l,r}N_a\,n_{a1}n_{a2}n_{a3}=16 &,&
T_i\;=\;\sum_{a=c,l,r}N_a\,n_{ai}m_{aj}m_{ak}=0,\;\;\;i\ne j\ne k\ne
i
\ea
where the indices $i,j,k$ take the values $1,2,3$ and refer  to the
three torii $T^2$.
 The additional constraints imposed on the winding numbers due to
the cancelation of mixed  anomalies  arise as
follows\cite{Ibanez:1998qp,Aldazabal:2000cn}. The gauge group
structure of the D-brane constructions involves symmetries
$U(N)_a=SU(N)_a\times U(1)_a$, thus mixed $SU(N)_b^2- U(1)_a$
anomalies appear, being of the form $A_{ab}=\frac
12\delta_{ab}\sum_cN_cI_{bc}+\frac 12\,N_bI_{ab}$. The first part is
proportional to the non-Abelian anomaly and therefore it is zero,
however the second part does not vanish even after imposing the
Tadpole conditions presented above. The cancelation of the $U(1)$
anomalies in these models is analogous to the heterotic case: closed
string modes coupled to gauge fields give rise to a generalized
GS-mechanism. In particular, in 10 dimensions there are RR-fields
$C_2,C_6$ which upon dimensional reduction give the
 two-form fields $ B_2^0=C_2$, $ B_2^i=\int_{T^{2}_j\times
T^{2}_k}C_6, i\neq j\neq k\neq i$ and their four-dimensional duals
$C^0=\int_{T^{2}_1\times T^{2}_2\times T^{2}_3}C_6$,
$C^i=\int_{T^{2}_i} C^{2}$. Moreover, the four-dimensional couplings
of the $B_2^i$ to the gauge fields are of the form $\sum_ic_a^i
B_2^i\wedge F_a$ and the $c_a^i$ depend on the winding numbers as
follows
\ba
N_am_{a1}m_{a2}m_{a3}\int_{M_4}B_2^0\wedge F_a &,&
N_an_{aj}n_{ak}m_{ai}\int_{M_4}B_2^i\wedge F_a\nonumber
\ea
Similarly, their dual scalars have couplings to the strengths $F_b$
of the form $d_b^j\,Tr[F_b\wedge F_b]$,
where\cite{Ibanez:1998qp,Aldazabal:2000cn,Ibanez:2001nd}
\ba
n_{b1}n_{b2}n_{b3}\int_{M_4}C^0\wedge F_b\wedge F_b &,&
n_{bi}n_{bj}n_{bk}\int_{M_4}C^i\wedge F_b\wedge F_b\nonumber
\ea
Combining both couplings, the anomalies $A_{ab}$ are canceled since
it holds
\ba
A_{ab}+\sum_i c_a^i d_b^i &=&0\label{anca}
\ea
Thus, to cancel the anomalies, some products $c_a^id_b^i$ should be
non-vanishing, while due to the $B_2^i$ couplings, some $U(1)$
combinations become massive. This also happens for the anomaly-free
$U(1)$'s whose corresponding combination does not couple to $F\wedge
F$. Notice that this mechanism does not involve any non-zero vev for
some scalar field, so as a result the $U(1)$'s  remain as global
symmetries in the low-energy effective theory.

We have already enumerated  the trinification spectrum which can in
principle be derived in the context of a D-brane configuration with
$U(3)^3$ gauge symmetry. Our intention here is to construct a
D-brane  non-supersymmetric analogue in the context of intersecting
branes, where supersymmetry is broken by gauge
fluxes~\cite{Bachas:1995ik,Angelantonj:2000hi} or in its dual
picture by the different intersection angles. Thus, in what follows
semi-realistic low energy effective trinification models with three
generations and the possible minimum number of other matter fields
are derived in the context of intersecting D-branes. More precisely,
Tadpole and anomaly cancelation conditions as well as other
constraints on the winding numbers of the branes around the three
torii are imposed to obtain classes of models with the required
effective field theory characteristics. We start with some mild
initial assumptions, and derive two particular families of solutions
which imply three generations and the minimum number of additional
fields.

 First, we will make the simplified assumption that the winding
number $n_{c2}$ of the color brane on the second torus is zero. A
simple inspection of the multiplicity equations shows that a
consistent solution could arise by setting two more $n_{ai}$'s equal
to zero. Thus, we start with
\ba
n_{c2}=n_{r1}=n_{l3}&=&0
\ea
We should note that these conditions immediately  imply that  the
contribution of the observable sector to the Tadpole condition $T_0$
is  zero, $\sum_iN_in_{i1}n_{i2}n_{i3}=0$. However, one can always
add extra branes along the orientifold plane without intersections
with the observable sector, whereas their contribution to $T_0$ can
be adjusted so that
\ba
T_0\;=\;\sum_aN'_an'_{a1}n'_{a2}n'_{a3}&=&16
\ea
We proceed further by imposing the necessary conditions to obtain
the desired spectrum. We first start with the sectors
$D_a-\Omega\,{\cal R}\,D_b$. We already noticed that the
representations ${\cal Q}'(3,3,1)$, and ${\cal Q}^{'c}(3,1,3)$
contain exotic quark fields, thus the corresponding quantities
\ba
I_{cl^*}&=& -m_{c2} m_{l3} n_{c3} n_{l2}\left(m_{l1} n_{c1}+m_{c1} n_{l1}\right)  \\
 I_{cr^*}&=&-m_{c2} m_{r1} n_{c1} n_{r2} \left(m_{r3} n_{c3}+m_{c3} n_{r3}\right)
 \ea
 representing their multiplicities in the spectrum should be zero. Hence, we
fist impose the conditions $I_{cl^*}=0$ and $I_{cr^*}=0$, which
imply the relations $ n_{c1}= -\frac{m_{c1} n_{l1}}{m_{l1}}$ and $
n_{r3}=-\frac{m_{r3}n_{c3}}{m_{c3}}$. There is one more
representation originating of the same  sector, namely ${\cal
L}'(1,3,3)$, which under the standard hypercharge definition
(\ref{HC}) contains fields with exotic charges. However, if one
extends the hypercharge definition according to (\ref{NHC}), the
fields obtain ordinary lepton charges. Their relevance in the
spectrum will be discussed in conjunction with the Tadpole
conditions.

Left- and right handed quarks are accommodated in ${\cal
Q^{\hphantom{c}}}=(3,\bar 3,1)$ and ${\cal Q}^c=(\bar 3,1,3)$ and
live in the intersection loci  $D_c-D_l$ and $D_c-D_r$ respectively,
while leptons are in ${\cal L}^{\hphantom{c}}=(1,3,\bar 3)$, and are
found in the intersection of $D_l-D_r$ brane stacks. The number of
generations is related to the intersections $I_{ab}$, thus we impose
the conditions $I_{cl}=3$, $I_{cr}=-3$ and $I_{lr}=3$ which lead to
further constraints on the winding numbers. Substituting in the
Tadpole conditions the $T_2=0$ is automatically satisfied, whilst
the remaining can be expressed as follows
\ba
T_1\;=\;m_{l2} m_{l3} n_{l1}\left(1- \frac{t}s\right) &,&
T_3\;=\;\frac{3}{ 2\,m_{l3} n_{l1} n_{l2}}\left(\frac{
s}{t}-1\right)\label{TCC}
\ea
where $r={m_{l1} m_{l2} m_{l3}}$  and $t={m_{c1} m_{c2} m_{c3}}$. In
addition the number $I_{lr^*}$ of the states $(1,3,3)$, is given by
\ba
I_{lr^*}&=&3\left(1-\frac st \right)\nonumber
\ea
The vanishing of the remaining Tadpole conditions (\ref{TCC})
require $r=t$, which also implies $I_{lr^*}=0$. Thus, the only
remaining fermionic states beyond those accommodating the three
families, arise from the ${\cal D}_a-\Omega{\cal R}{\cal D}_a $
sector.  The antisymmetric representations are
\ba
I_{aa^*}&=& m_{a1} m_{a2} m_{a3}\;=\;8 t,\label{AR}
\ea
while there are also symmetric as well as antisymmetric ones whose
multiplicity is given by
\ba
I'_{aa^*}&=&4(n_{a1} n_{a2} n_{a3}-1) m_{a1} m_{a2}
m_{a3}\;=\;-4t\label{SAR}
\ea
Thus,  there is an equal number of  antisymmetric representations
for each gauge group factor $a=C,L,R$, and this  also holds for the
symmetric ones. It is worth noticing that the result $I'_{aa^*}=-4
t$ imposes one more constraint on the winding numbers $m_{ai}$.
Indeed, since in the present case $ m_{ai}\ne 0$ and $ t=m_{a1}
m_{a2} m_{a3}$ while $4t$ should be an integer, we conclude that at
most two torii can be titled.
 Summarizing, there is a family of solutions of three generation
 trinification models determined by the following relations imposed
 on the wrapping numbers $(n_{ai},m_{ai})$,
\ba
m_{r2}&=& -\frac{m_{l2} n_{r2}}{n_{l2}}\nonumber
\\
m_{r3}&=&-\frac{t\, n_{l2}}{m_{l2} m_{r1} n_{r2}}\nonumber
\\
n_{r3}&=&\frac{3m_{l1}}{2 t\, m_{r1} n_{l1} n_{r2}}\label{fincond}
\\
n_{c1}&=&-\frac{m_{c1}  n_{l1}}{m_{l1}}\nonumber
\\
n_{c3}&=&\frac{3m_{c3}}{2 t\, m_{l3} n_{l1} n_{l2}}\nonumber
\ea
One can be re-express the left-hand side of the above expressions in
terms of seven integer parameters and seek explicit solutions. In
Table \ref{M3} a specific example of winding numbers satisfying the
above constraints is presented. They imply three generations and the
existence of $I_{aa^*}=-2$ antisymmetric as well as $I'_{aa^*}=1$
symmetric representations for each group factor.
\begin{table}[!t]
\centering
\renewcommand{\arraystretch}{1.2}
\begin{tabular}{|c|c|c|c|}
\hline
$ N_a$ &$(n_{a1},m_{a1})$& $(n_{a2},m_{a2})$& $(n_{a3},m_{a3})$\\
\hline
 $N_c=3$&   $(2,\frac 12)$& $(0,-1)$ & $(-3,\frac 12)$\\
\hline
 $N_l=3$&  $ (-2,\frac 12)$&$ (-1,-1)$  & $(0,\frac 12)$\\
    \hline
$ N_r=3$& $(0,\frac 12)$ & $(1,-1)$ &$(3,\frac 12)$ \\
\hline
\end{tabular}
\caption{\label{M3}A specific example of winding numbers for the
case (\ref{fincond}). }
\end{table}

We have already asserted that in the D-brane construction of the
trinification model one is confronted with the appearance of mixed
$U(1)_a-SU(3)_b$ anomalies $A_{ab}$ due to the three extra abelian
factors. As discussed, two linear $U(1)_{C,L,R}$ combinations have
mixed anomalies with the non-abelian gauge symmetries while the
linear combination (\ref{aas}) is anomaly free. According to our
previous discussion, the anomalies in the present model are canceled
by a generalized Green-Schwartz mechanism. In particular, the
non-zero coefficients $c_a^i$  involved in the anomaly cancelation
conditions (\ref{anca}) are found to be
\ba
c_r^1:&B_2^1\wedge (3 n_{r2}n_{r3}m_{r1})\,F_r \nonumber
\\
c_c^2:&B_2^2\wedge (3 n_{c3}n_{c1}m_{c2})\,F_c \nonumber
\\
c_l^3:&B_2^3\wedge (3 n_{l1}n_{l2}m_{l3}) \,F_l\nonumber
\\
c^0:&B_2^0\wedge \,t\,(F_c+F_l+F_r)\nonumber
\ea
The coefficients $d_a^i$ can be calculated similarly. Using the
relations (\ref{fincond})  one finds that the sums in (\ref{anca})
are
$$
\sum_{i=1}^3 c_l^id_r^i\;= \;\sum_{i=1}^3
c_c^id_l^i\;=\;-\sum_{i=1}^3 c_c^id_r^i\;=\;-\frac 92,
$$
which consistently cancel  the triangle mixed anomalies given by
$A_{ab}=\frac 12 N_b\,I_{ab}$.

It it to be noticed that in  this particular solution the $B_2^0$
field couples to the linear combination
$U(1)_{Z'}=U(1)_C+U(1)_L+U(1)_R$, since
$c_a^0=t=m_{a1}m_{a2}m_{a3}\ne 0$ and $\sum_ac_a^0\ne 0$.\footnote{
In its most general definition (\ref{NHC}) the hypercharge is a
linear combination of the $X_{L',R'}$ generators of the
corresponding non-abelian gauge groups and the three factors
$U(1)_{C,L,R}$. As far as the component $Y_{Z'}=\sum_{a=C,L,R} Q_a$,
related to the three abelian factors is concerned, we need to ensure
that no closed string mode renders it massive, thus we should have
$\sum_{a=C,L,R} \,c_a^i=0,\;\forall\,i$. This, however, cannot be
satisfied in the present example.  }  Therefore the anomaly -free
$U(1)_{Z'}$ becomes also massive while  the corresponding abelian
symmetry is broken and survives only as a global symmetry in the
low-energy effective model. Clearly, in this case the hypercharge
redefinition (\ref{NHC}) is not possible. This fact has rather
significant impact on the properties of the additional spectrum of
the model, since under the conventional definition of the
hypercharge, the antisymmetric triplets carry exotic fractional
electric charges.

Perhaps, the most annoying exotic representations in the previous
example are those having non-trivial transformation properties under
the color gauge group, namely the symmetric and antisymmetric
representations originated from open strings with both end-points on
$U(3)_c$.  It is possible to eliminate those states by demanding
$I_{cc^*}=0$. For example, starting with the condition $m_{c2}=0$,
one immediately obtains $I_{cc^*}= m_{c1}m_{c2}m_{c3}=0$. Proceeding
as in the previous case, we may further impose the conditions to
obtain three fermion generations and $I_{cl^*}=I_{cr^*}=0$ to avoid
additional exotic states. To satisfy Tadpole conditions we need to
impose further constraints. For example, eliminating $T_1$, we solve
for $m_{c1}$ and substitute to the remaining conditions. In this
case we find also that $T_3=0$ is simultaneously satisfied, while
\ba
T_2\;=\;\frac{12 x y (2 x y+3)^2 m_{r1}}{(4 x y+3)^2 m_{l1} m_{l2}
n_{l3} n_{r1}}&=&0 \label{t40}\\ T_0\;=\;\frac{24 x^2 y (2 x y+3)
n_{c3}}{(4 x y+3)^2 n_{l3}}&=&16 \label{t4A}
\ea
where $ y= m_{c3} m_{l1} m_{l2}$ and $x= n_{c1} n_{c2} n_{l3}$.
Apparently the condition (\ref{t40})  can be satisfied by setting
$m_{r1}=0$, but a simple analysis shows that we should also have
$n_{l3}=0$ which further implies $T_0=0$. Hence, condition
(\ref{t4A}) can be fulfilled only in the presence of hidden matter
in the sense described also in the previous case. It can be checked
that in this case the only extra representations remaining in the
spectrum are the $SU(3)_L$ antisymmetric ($A_L$) and symmetric
($S_L$) ones with multiplicities $\pm\,6/z $ and $-6/z $
respectively, where $z$ is the integer   $z=n_{c1} n_{c2} n_{c3}$.
Summarizing, the conditions imposed on the winding numbers for this
particular solution, are
\ba
n_{l1}\;=\; -\frac{m_{l1} n_{c1}}{m_{c1}}&,&m_{l3}\;=\;
   \frac{3}{2 m_{l1} m_{l2} n_{c1} n_{c2} n_{c3}}\nonumber
\\
n_{r2}\;=\; \frac{m_{c3} n_{c1} n_{c2}}{m_{r3} n_{r1}}&,&m_{r2}\;=\;
\frac{3}{2 m_{c1} m_{r3} n_{c2} n_{c3}
   n_{r1}}\label{sol2}
\\
n_{r3}\;=\;  -\frac{m_{r3} n_{c3}}{m_{c3}}&,&n_{l2}\;=\;
-\frac{2}{3} m_{c1} m_{c3} m_{l2} n_{c1} n_{c2}^2 n_{c3}\nonumber
\ea
with  $|z|$ assuming one of the integer values $1,2,3,6$. In Table
\ref{M4} a specific example is given for this second case.
\begin{table}[!t]
\centering
\renewcommand{\arraystretch}{1.2}
\begin{tabular}{|c|c|c|c|}
\hline
$ N_a$ &$(n_{a1},m_{a1})$& $(n_{a2},m_{a2})$& $(n_{a3},m_{a3})$\\
\hline
 $N_c=3$&   $(1,\frac 12)$& $(2,0)$ & $(3,\frac 12)$\\
\hline
 $N_l=3$&  $ (-1,\frac 12)$&$ (-1,\frac 12)$  & $(0,1)$\\
    \hline
$ N_r=3$& $(1,0)$ & $(2,1)$ &$(-3,\frac 12)$ \\
\hline
\end{tabular}
\caption{\label{M4}A specific example of winding numbers for
(\ref{sol2}). }
\end{table}

We concentrate now on some aspects of the low energy effective model
implied by the above construction. A decisive factor for the
viability of a model is its prediction for the fermion mass
spectrum. Indeed, among the unresolved problems within the Standard
Model theory, is the hierarchy of the fermion masses. The revelation
of a mechanism generating such a hierarchy would be an attractive
feature of any model incorporating the SM theory. In intersecting
D-brane models Yukawa couplings are suppressed by an exponential
factor $\lambda_{ijk}\sim e^{-A_{ijk}}$, where $A_{ijk}$ is the
triangle world-sheet area whose vertices are identified at the
locations of the fields participating in the coupling.  It is
expected that this geometrical nature of the couplings is intimately
related to the hierarchical nature of the fermion masses. Another
important role in the determination of the Yukawa potential is
played by the various surplus $U(1)$ factors which    survive as
global symmetries at low energies. In the present model in
particular, due to the existence of the additional $U(1)_{C,L,R}$
symmetries, only the following Yukawa coupling is present  at the
tree-level Yukawa potential
\ba
\lambda_{ijk}^{q}{\cal Q}_i\,{\cal Q}_j^c\,{\cal H}_k, \;\;{k=1,2,3}
\ea
providing up and down quark masses as well as masses for the extra
triplets.  It should be noticed that in these constructions scalar
fields with the same quantum numbers and multiplicity as the
corresponding fermions are found at the intersections.  The fields
$H_a=(1,3,\bar 3)$ participating in the above coupling carry the
same quantum numbers with leptons and have the same multiplicity.
This is a welcome fact since in the trinification model at least two
Higgs are needed to obtain realistic quark mixing.

On the contrary, the extra $U(1)_{C,L,R}$ factors do not allow the
existence of tree-level lepton couplings in the Yukawa potential.
The lowest order allowed leptonic Yukawa terms arise at fourth
order,
\ba
\frac{f_{ij}^{ab}}{M_S}\, {\cal H}_a^\dagger \,{\cal
H}_b^\dagger\,{\cal L}_i\,{\cal L}_j
\ea
where $M_S$ is the effective string scale. These terms  induce
masses to the charged leptons suppressed by a factor $M_R/M_S$
compared to quark masses, providing thus, a natural quark-lepton
hierarchy. They also predict Dirac and Majorana neutrino mass
terms\cite{Leontaris:2005ax}. However, additional terms should be
considered in order to provide with masses the exotic matter fields.
The perturbative  Yukawa terms considered above are not the only
source of fermion masses  in string theory. Indeed, it has been
known some time ago that the space-time superpotential, is modified
by world-sheet instantons\cite{Dine:1986zy}, while recently
mechanisms through instanton effects for generation of masses and
the decoupling of the exotic matter have been
examined~\cite{Blumenhagen:2006xt,Ibanez:2006da,Blumenhagen:2007zk}.
A more detailed discussion for the fermion mass problem and other
 issues of vital importance in this model will be devoted in a longer
publication.

\end{document}